\documentclass[doublecol]{epl2}

\title{Transport properties in  quantum Griffiths phases}
\shorttitle{Transport properties in quantum Griffiths phases} 

\author{David Nozadze and  Thomas Vojta}
\shortauthor{D. Nozadze and  T. Vojta}

\institute{
 Department of Physics, Missouri University of Science $\&$ Technology, Rolla, MO 65409
}



\abstract{
We study the electrical resistivity in  the quantum Griffiths phase associated  with the antiferromagnetic quantum phase transition in  a metal.
The resistivity is calculated by means of  the semi-classical Boltzmann equation. We  show that the scattering of electrons by locally ordered rare
regions leads to a singular temperature dependence. The rare-region contribution to the resistivity  varies as $T^\lambda$ with temperature $T,$ where    $\lambda$ is the usual Griffiths exponent which takes the value  zero at the critical point and increases with  distance from  criticality. We find similar singular contributions to other transport properties such as thermal resistivity, thermopower and the Peltier coefficient. We also compare our results with existing  experimental data and suggest new experiments.
}

\begin{document}

\maketitle

\section{Introduction}

Quantum phase transitions \cite{Sachdevbook}  occur at zero temperature when an external parameter such as magnetic field, pressure or chemical composition
is varied. They are driven by  quantum rather than thermal fluctuations. At continuous quantum phase transitions, i.e., quantum critical points,
the quantum  fluctuations driving the transition  diverge  and become scale invariant in space and time. These fluctuations  dominate the material's properties in the vicinity of the quantum critical point at low but non-zero temperatures. In metallic systems, they can
cause strong deviations from the conventional Fermi-Liquid behavior of normal metals \cite{1994_Lohneysen_PRL}.

Impurities, defects or other kinds of quenched disorder can significantly modify the low temperature behavior of quantum many-particle
systems. The interplay between dynamic quantum fluctuations and static disorder  fluctuations leads to much more dramatic effects at quantum phase transitions  than at classical thermal
phase transitions, including quantum Griffiths singularities \cite{1969_Griffiths_PRL, 1995_Thill_Physyca_A, 1996_Rieger}, infinite randomness
critical points featuring exponential instead of power-law scaling \cite{1992_Fisher_PRL, 1995_Fisher_PRB} and the smearing of
the phase transition \cite{2003_Vojta_PRL}. These unconventional phenomena are   caused by large spatial
regions (rare regions) that are devoid of impurities and can show local order even if the bulk system is in
the disordered phase. The fluctuations of these rare regions are very slow because they require changing the
order parameter in a large volume. Griffiths showed that this leads to a singular free energy in a whole
parameter region which is now known as the Griffiths phase.  The probability $\mathcal{P}(L^d)$  for finding
an impurity-free rare region with  linear size $L$ in a disordered system is exponentially small in its volume $L^d$,
$\mathcal{P}(L^d)\sim \exp(-cL^d)$ with $c$ being a constant that depends on the disorder strength. In
systems in which the characteristic energy  $\epsilon$ of such a rare region decays exponentially
with its volume, $\epsilon\sim\exp(-bL^d),$ the resulting  density of states is of power-law type,
$\rho(\epsilon)\propto\epsilon^{\lambda-1},$
where $\lambda=c/b$ is the non-universal Griffiths exponent. It varies systematically  within the
Griffiths phase and vanishes at the critical
point. The power-law density of states  $\rho(\epsilon)$ leads to non-universal   power-law quantum Griffiths
singularities  of several  thermodynamical  observables including the specific heat, $C\sim T^\lambda,$ and the magnetic
susceptibility, $\chi\sim T^{\lambda-1}.$  The zero-temperature magnetization-field curve
behaves as $M\sim H^{\lambda}$ (for reviews, see Refs. \cite{2006_Vojta_JPhysA, 2010_Vojta_JLTPhys}).

Quantum Griffiths phases have been predicted to occur  not only in localized magnets but also
in metallic systems  \cite{2000_Castro_PRB, 2007_Hoyos_PRL, 2005_Vojta_PRB}, but clear-cut experimental verifications have been
absent for a long time. Only recently, quantum  Griffiths phases have been observed in
  experiment in a number of systems such as
magnetic semiconductors \cite{2007_Guo_PRL, 2010_Guo_PRB1, 2010_Guo_PRB2}, Kondo lattice ferromagnets
 \cite{2007_Sereni_PRB, 2009_Westerkamp_PRL}  and transition metal ferromagnets \cite{2010_Sara_PRL}.
The  lack of experimental evidence for quantum Griffiths phases in metals may be (at least partially) due to the theories
 being incomplete: while the thermodynamics in quantum Griffiths phases is comparatively well understood, very little is known about the experimentally
important and easily accessible   transport  properties.

In this Letter we therefore study the electrical resistivity in the quantum Griffiths phase of an antiferromagnetic metal  by means of the
semi-classical Boltzmann equation approach. In the same manner, we also investigate  other transport properties
such as the thermal resistivity, the thermopower and the Peltier coefficient. We find that the scattering of the  electrons  by
spin-fluctuations in the rare regions leads to singular temperature dependencies not just at the quantum critical point but
in the entire antiferromagnetic quantum Griffiths phase. The rare region contribution to the resistivity varies as
$\Delta \rho\propto T^{\lambda}$
with temperature $T$,
the contribution to thermal resistivity behaves as $\Delta  W  \propto T^{\lambda-1}$, and   the thermopower
and  the Peltier coefficient behave as $\Delta S  \propto \ T^{\lambda+1}$ and $ \Delta \Pi \propto \ T^{\lambda+2},$ respectively.

\section{Model and method of solution}

Let us now sketch the derivation of these results.
The transport properties of the itinerant  antiferromagnetic systems we are interested in can be described by
a two-band model consisting of $s$ and $d$ electrons \cite{1966_Mills_JPhys, 1975_Ueda_JPhysSJ}. The Hamiltonian has the form
$H=H_{s}+H_{d}+H_{s-d},$ where $H_{s}$ and  $H_{d}$ are the Hamiltonians of  $s$ and $d$ electrons, respectively.
$H_{s-d}$ corresponds to the exchange interaction between  $s$ and $d$ electrons.
Only the $s$ electrons contribute to the transport properties. They are scattered
by the spin-fluctuations of the $d$ electrons which are assumed to be in the antiferromagnetic quantum
Griffiths phase. The contribution to the  resistivity  from the scattering by the spin-fluctuations stems from
the $s-d$ exchange interaction term of the Hamiltonian
\begin{eqnarray}
H_{s-d}=g \int d  \textbf{r} \ \textbf{s}(\textbf{r})\cdot \textbf{S}(\textbf{r})\,,
\end{eqnarray}
where $g$ is the  coupling between $s$ and $d$ electrons. \textbf{s} and
\textbf{S} are the  spin densities of the $s$ and $d$ electrons, respectively.

Close to an antiferromagnetic transition in  three-dimensional space, transport properties can be treated within a semi-classical approach using the Boltzmann
equation because quasiparticles are still (marginally) well defined. For simplicity,  we  also assume
that the spin-fluctuations are in equilibrium, i.e., we neglect drag effects.
This approximation is valid if the system can lose momentum efficiently by
Umklapp or impurity scattering as is the case in a dirty antiferromagnetic system.
The linearized Boltzmann equation in the presence of an electric field $\textbf{E}$ and
a temperature gradient $\nabla T,$ but zero magnetic field can be written as \cite{Zimanbook}
\begin{eqnarray}
-\textbf{v}_\textbf{k}\frac{\partial f^0_\textbf{k}}{\partial T} \nabla T
-\textbf{v}_\textbf{k}\frac{\partial f^0_\textbf{k}}{\partial \varepsilon_\textbf{k}}
 \textbf{E} =\biggl(\frac{\partial f_\textbf{k}}{\partial t}\biggl)_{scatt}\,,
\end{eqnarray}
where $f^0_\textbf{k}$  is the equilibrium Fermi-Dirac distribution function.
The first and second  terms correspond to the rate changes of the electron distribution
function $f_\textbf{k}$ due to the diffusion and electric field $\textbf{E}$, respectively.
The last one is the collision  term.
Let the stationary solution of the Boltzmann equation be $f_\textbf{k}=f^0_\textbf{k}-\Phi_\textbf{k}
(\partial f^0_\textbf{k}/\partial \varepsilon_\textbf{k}),$
where $\Phi_\textbf{k}$ is a measure of the deviation  of the electron distribution from equilibrium. Then the linearized
scattering term due to the spin-fluctuations  has the form \cite{1995_Hlubina_PRB, 1975_Ueda_JPhysSJ}

\begin{eqnarray}
\biggl(\frac{\partial f_\textbf{k}}{\partial t}\biggl)_{scatt}&=&\frac{2g^2}{T}\sum_{\textbf{k}'}f^0_{\textbf{k}'}(1-f^0_\textbf{k})n(\varepsilon_\textbf{k}-\varepsilon_{\textbf{k}'}) \nonumber \\
&\times&  \rm{Im} \chi(\textbf{k}-\textbf{k}', \varepsilon_\textbf{k}-\varepsilon_{\textbf{k}'})(\Phi_\textbf{k}-\Phi_{\textbf{k}'})\,,
 \nonumber \\ &=& \frac{1}{T}\sum_{\textbf{k}'} \mathcal{P}_{\textbf{k}'}(\varepsilon_\textbf{k}-\varepsilon_{\textbf{k}'})(\Phi_\textbf{k}-\Phi_{\textbf{k}'})
\end{eqnarray}
where $n(\varepsilon_\textbf{k}-\varepsilon_{\textbf{k}'})$ is the Bose-Einstein distribution function and
$\chi$ is the total dynamical susceptibility of the spin-fluctuations of the $d$ electrons.

\section{Electrical resistivity}

In order  to calculate the electrical  resistivity we  consider Ziman's variational principle \cite{Zimanbook}.
The resistivity $\rho$  is given as the minimum of a functional of $\Phi_\textbf{k}$  \cite{Zimanbook}
\footnote{We set Plank's constant, electron's  charge and Boltzmann constant $\hbar=e=k_{B}=1$ in what follows.}
\begin{eqnarray}
\rho[\Phi_\textbf{k}]=\rm{min}
\Biggl[\frac{1}{2T}\frac{\int\int(\Phi_\textbf{k}-\Phi_{\textbf{k}'})^2\Gamma^{\textbf{k}'}_{\textbf{k}}d\textbf{k} d\textbf{k}'}{\bigl(\int v_\textbf{k} \Phi_\textbf{k}\frac{\partial f^0_\textbf{k}}{\partial \varepsilon_\textbf{k}} d\textbf{k}\bigr)^2}\Biggr]\,,
\end{eqnarray}
where
\begin{eqnarray}
\Gamma^{\textbf{k}'}_{\textbf{k}}=
\int_{0}^{\infty}d  \omega\  \mathcal{P}_{\mathbf{k}'} (\omega)\delta(\varepsilon_\mathbf{k'}-\varepsilon_{\mathbf{k}}+\omega) \,.
\end{eqnarray}

Quantum Griffiths effects in disordered metallic systems  are realized both in Heisenberg magnets \cite{2005_Vojta_PRB,  2007_Guo_PRL} and
in Ising magnets. In the latter case, they occur in a transient temperature range where the damping is unimportant \cite{2000_Castro_PRB}.
In the following, we consider both cases.

As we are interested in the rare-region contribution to the resistivity in the Griffiths phase, we need to find the rare region dynamical susceptibility
which is simply the sum over the susceptibilities of the individual rare regions.
The imaginary part of the  dynamical  susceptibility of a single cluster (rare region) of characteristic energy $\epsilon$
in the  quantum Griffiths phase of a disordered itinerant
quantum Heisenberg  antiferromagnet is given by
\begin{eqnarray}
\rm{Im} \chi_{cl}    (\mathbf{q},\omega; \epsilon )=\frac{\mu^2 \gamma \omega}{\epsilon^2(T)
+\gamma^2\omega^2}F_{\epsilon}^2(\mathbf{q})\,,
\end{eqnarray}
where $\mu$  is the moment of the cluster and $\gamma$ is the damping coefficient which results from the coupling of the spin-fluctuations
and the electrons.
 $\epsilon(T)$ plays the role of the local  distance from criticality. For high temperatures $\gamma T \gg \epsilon,$ $\epsilon(T)\approx T$ and for low temperatures
$\gamma T \ll \epsilon,$ $\epsilon(T)\approx \epsilon.$ $F_{\epsilon}(\mathbf{q})$ is the form factor of the cluster
which encodes the spatial magnetization profile.
For random quantum Ising models the imaginary part of the  dynamical magnetic susceptibility of a single cluster (rare region) is given by
 \begin{eqnarray}
\rm{Im} \chi_{cl}(\mathbf{q},\omega; \epsilon )&=&\pi \frac{\mu^2}{4}\tanh \left (\frac{\epsilon} {2T} \right ) \nonumber \\ &\times & [\delta (\epsilon-\omega)- \delta (\epsilon+\omega)]F_{\epsilon}^2(\mathbf{q})\,.
\end{eqnarray}

 To get the total rare-region susceptibility, we integrate over all rare regions using the  density
of states $\rho(\epsilon),$
\begin{eqnarray}
\rm{Im} \chi(\mathbf{q},\omega )=\int_{0}^{\Lambda}d\epsilon \rho(\epsilon)  \rm{Im} \chi_{cl}(\mathbf{q},\omega; \epsilon)\,,
\end{eqnarray}
where $\Lambda$ is an energy cut-off. The precise functional form of $F_{\epsilon}(\mathbf{q})$
is not known, since every cluster has a different shape and size. However, we can  find it approximately by
analyzing the Fourier transform of a typical  local magnetization profile of the rare region. Consider a rare region of linear size $L$
(located at the origin).
Following    Millis \emph{et al.} \cite{2001_MIllis_PRL}, the
order parameter is approximately  constant for  $r<L,$  while for  large $r>L,$ it decays as $e^{-r/\xi}/r,$
where $\xi$ is the bulk correlation length.
Taking the Fourier transform we find  that  $F_{\epsilon}(\mathbf{q})$  depends on $\varepsilon$  via the combination $|\mathbf{Q}-\mathbf{q}|^{3} \log (\epsilon^{-1})$ only, where $\mathbf{Q}$ is the ordering wave vector.
Correspondingly, from Eq.(8), we find that the rare region contribution to the zero-temperature susceptibility in the quantum Griffiths phase
 can be expressed  as
\begin{eqnarray}
\rm{Im} \chi(\mathbf{q},\omega) \propto  |\omega|^{\lambda-1} \rm{sgn}(\omega)\, X[(\mathbf{q}-\mathbf{Q})^3\log (\omega^{-1})] \,,
\end{eqnarray}
where $X$ is a scaling function. The precise form of the logarithmic correction is difficult to
find and beyond the  scope of this paper. For random quantum Ising models, the susceptibility has the same structure as Eq.(9) \cite{2000_Castro_PRB}.  It is clear that the scaling function $X$ will give only logarithmic corrections to the temperature dependence of the resistivity $\rho$ in our further calculations.

To minimize the resistivity functional (4), we need to make an ansatz for the distribution $\Phi.$
Close to an antiferromagnetic quantum phase transition, the magnetic scattering is highly anisotropic
because  $\chi(\mathbf{q},\omega)$ peaks around the ordering wave vector $\mathbf{Q}.$ However,  since we are interested
in a strongly disordered system, the low-temperature  resistivity will be dominated by the elastic impurity  scattering
which is isotropic  and redistributes the electrons over the Fermi surface. Correspondingly, we can use the standard ansatz
 \begin{eqnarray}
\Phi_{\mathbf{k}} \propto \textbf{n}\cdot\textbf{k}\,.
\end{eqnarray}
 where $\textbf{n}$ is a unit vector parallel to the electric filed. Note that any constant prefactor in $\Phi_{\textbf{k}}$ is unimportant because it drops out
 in the resistivity functional (4) and in the corresponding thermal resistivity functional (13).
  Then, after applying standard techniques \cite{Zimanbook} the magnetic  part of the resistivity  given in  Eq.\ (4) becomes
\begin{eqnarray}
\Delta \rho \propto T \int d^3\textbf{q} \frac{(\textbf{n}\cdot \textbf{q})^2}{q} \int_{0}^{\infty}
d\omega \frac{\partial n(\omega)}{\partial T} \rm{Im} \chi(\textbf{q},\omega)\,.
\end{eqnarray}

Inserting the susceptibility (9) yields the rare-region contribution to the resistivity
in the    antiferromagnetic quantum Griffiths phase as
\begin{eqnarray}
\Delta \rho  \propto  T^\lambda \,.
\end{eqnarray}
Thus, the temperature-dependence of the resistivity follows a non-universal power-law governed by the Griffiths exponent $\lambda$.

\section{Other transport properties}

In the same way, we study other transport properties such as the thermal resistivity, the thermopower, and the Peltier coefficient.
The variational principle for the thermal resistivity has the form \cite{Zimanbook}
\begin{eqnarray}
W[\Phi_\textbf{k}]=\rm{min}\Biggl[\frac{\int\int(\Phi_\textbf{k}-\Phi_{\textbf{k}'})^2\Gamma^{\textbf{k}'}_{\textbf{k}}d\textbf{k} d\textbf{k}'}{\bigl(\int v_\textbf{k} (\varepsilon_\textbf{k}-\mu)\Phi_\textbf{k}\frac{\partial f^0_\textbf{k}}{\partial \varepsilon_\textbf{k}} d\textbf{k}\bigr)^2}\Biggr]\,,
\end{eqnarray}
where $\mu$ is the chemical potential of the s-electrons.
As long as impurity scattering dominates, we can use the standard  ansatz for the variational function,
\begin{eqnarray}
\Phi_{\mathbf{k}} \propto (\varepsilon_\textbf{k}-\mu)\mathbf{n}\cdot\textbf{k}\,.
\end{eqnarray}
Then, following the calculation for the thermal resistivity outlined in Ref. \cite{Zimanbook}
we obtain
\begin{eqnarray}
\Delta W& \propto & \frac{1}{T^2} \int d^3\textbf{q} \int d\omega \frac{\partial n(\omega)}{\partial T} \rm{Im} \chi(\textbf{q},\omega)
 \\  \nonumber &\times& \left[\omega^2\left( \frac{1}{q}-\frac{q}{6{k_{F}}^2}\right) +\frac{\pi^2q}{3{k_{F}}^2} T^2 \right]\,.
\end{eqnarray}
where $k_{F}$ is Fermi momentum of the $s$-electrons
\footnote{Here, we have averaged  over all directions of the vector $\mathbf{n}$; this is  sufficient to get
the temperature dependence.}.
Inserting the susceptibility (9) into (15), the temperature dependence of the thermal resistivity due to the spin-fluctuations in the Griffiths phase from the above equation  is given by
\begin{eqnarray}
\Delta W\propto T^{\lambda-1} \,.
\end{eqnarray}

The existence of an electric field $\mathbf{E}$ in a metal subject to a thermal gradient $\nabla T$
is called Seebeck effect and  is characterized  by the thermopower $S$ which is defined via $\mathbf{E}=S \ \nabla T $. To calculate the thermopower, we analyze the Boltzmann equation (2)
in the presence of both $\mathbf{E}$ and $\nabla T$ using the trial
function
\begin{eqnarray}
\phi_{\textbf{k}} \propto \eta_1 \textbf{n} \cdot \textbf{k}+\eta_2 (\varepsilon_\textbf{k}-\mu)\textbf{n}\cdot\textbf{k}\,.
\end{eqnarray}
where $\eta_{1}$ and $\eta_{2}$ are variational parameters. Elastic
impurity scattering leads to the usual linear temperature dependence $S_{imp}\propto T$ while the contribution due to the magnetic scattering
by the rare regions in the Griffiths phase reads
\begin{eqnarray}
\Delta S\propto   T^{\lambda+1} \,.
\end{eqnarray}

Another transport coefficient called the Peltier coefficient $\Pi$ characterizes the flow of
a thermal current in a metal in the absence of a thermal gradient. It is related to the thermopower
by $\Pi=S\ T.$ Correspondingly, the rare-region contribution to the Peltier coefficient has the form
\begin{eqnarray}
\Delta \Pi \propto \ T^{\lambda+2} \,.
\end{eqnarray}

\section{Discussion and conclusions}

In summary, we have investigated the transport properties in the quantum
Griffiths phase close to an antiferromagnetic quantum phase transition
in a metallic system (see Fig.\ 1).
\begin{figure} [t]
\centering
\includegraphics[width=7cm]{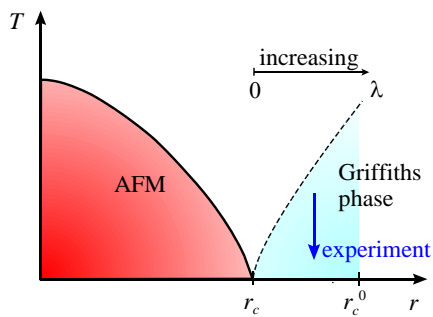}
\caption{(Color online) Schematic temperature-control parameter
   phase diagram of an itinerant antiferromagnet  close to the quantum
   critical point. Our results apply in the Griffiths phase at low temperatures.}
\end{figure}
The rare-region contributions to electrical resistivity, thermal resistivity,
thermopower, and the Peltier coefficient are characterized by non-universal
power-laws in $T$ which are controlled  by the Griffiths exponent $\lambda.$

Our results have been obtained using the semi-classical Boltzmann equation approach. This approach
is valid in Griffiths phase in which the system consists of a few locally ordered  rare regions in
a non-magnetic bulk where the quasiparticles are well-defined. Sufficiently close to the
actual quantum critical point (which is of infinite-randomness type) the quasiparticle description
may break down, invalidating our results. A detailed analysis of this question hinges on
the fate of the fermionic degrees of freedom  at the infinite-randomness quantum critical point.  This difficult
problem remains a task for the future.

We have used the standard isotropic ansatz (10,14) for the deviation of the electron distribution
from equilibrium.  This is justified as long as  the rare-region part $\Delta \rho (T)$  of the
resistivity is small compared to the impurity part $\rho_0.$  When $\Delta \rho$ becomes
larger, the anisotropy of the scattering needs to be taken into account. This can be done
by adapting  the methods of Rosch \cite{1999_Rosch_PRL} to the situation at hand.

We emphasize that our results have been derived for
antiferromagnetic quantum Griffiths phases and may not be valid for ferromagnetic systems. The problem is that
a complete theory of the ferromagnetic quantum Griffiths phase in a metal does not exist. In particular, the
dynamical susceptibility still is not known.
Correspondingly, the transport properties in  ferromagnetic quantum  Griffiths phases
remain an open problem.

Non-universal power-laws in a variety of observables including transport properties
can also arise from a different physical mechanism far away from the magnetic quantum
phase transition.  In   Kondo-disordered  systems, the existence of a wide distribution
of local single-ion  Kondo temperatures is assumed, this leads to the power-law singularities \cite{1997_Miranda_PRL,1996_Miranda_JPhysCond}.
This model was used to explain experimental results in some heavy fermion compounds such as $\rm{UCu_{4}Pd}$ and $\rm{UCu_{3.5}Pd_{1.5}}$
\cite{1995_Bernal_PRL,1996_Aronson_JPhysCond}.

Let us now turn to experiment. Unfortunately and somewhat ironically, all clear-cut experimental observations of
quantum Griffiths phases are  in itinerant ferromagnets rather than in antiferromagnets.
However, quantum Griffiths effects have been discussed  in the context  of the antiferromagnetic
quantum phase transition in heavy fermion systems \cite{2001_Stewart_RMP, 2006_Stewart_RMP}. One of the most striking  predictions following from our theory
is that the exponent  characterizing the electrical resistivity should be less than one
sufficiently close to the quantum phase transition.  There are several antiferromagnetic systems
such as $\rm{CeCo_{1.2}Cu_{0.8}Ge_{2}}$ and $\rm{Ce(Ru_{0.6}Rh_{0.4})_{2}Si_2}$ \cite{2001_Stewart_RMP, 2006_Stewart_RMP} that show
unusual power-law behaviour of the electrical resistivity  with an exponent less than unity. The
first system's resistivity increases with decreasing temperature. This is incompatible
with our prediction and described by the Kondo model. The resistivity of the second compound
decreases with decreasing temperature in agreement with our prediction. However, it is not clear
whether this behaviour is indeed  caused by the quantum Griffiths phase. To establish this, one should measure various thermodynamics
quantities as well as the transport properties  and relate their low-temperature behavior.

\acknowledgments
This work has been supported by the NSF under Grant No. DMR-0906566.

\bibliographystyle{eplbib}

\end{document}